\documentclass[preprint,12pt]{aastex}
\def \m3{{\rm Mark III}}

\def \<{\langle}
\def \>{\rangle}

\def\gsim{~\rlap{$>$}{\lower 1.0ex\hbox{$\sim$}}}
\def\lsim{~\rlap{$<$}{\lower 1.0ex\hbox{$\sim$}}}

\begin{document}

\title{Dark halos and elliptical galaxies as marginally stable dynamical systems}
\author{A. A. El Zant}
\affil{Centre for Theoretical Physics, Zewail City of Science and Technology, Sheikh Zayed, 12588, Giza, Egypt and\\
The British University in Egypt,Sherouk City, Cairo 11837, Egypt}%
\date{\today}
\begin{abstract}
The origin of equilibrium gravitational configurations is sought in terms of the stability of their trajectories, as described by the curvature of their Lagrangian configuration manifold of particle positions --- a context in which subtle spurious effects originating from the singularity in the two body potential become particularly clear.  We focus on the  case of spherical systems, which support only regular orbits in the collisionless limit, despite the  persistence of  local  exponential instability of $N$-body trajectories in the anomalous case of discrete point particle representation even as $N \rightarrow \infty$. When the singularity in the potential is removed, this apparent contradiction disappears. In the absence of fluctuations, equilibrium configurations generally correspond to positive scalar curvature, and thus support stable trajectories.   A null scalar curvature is associated with
an effective, averaged, equation of state describing dynamically relaxed equilibria with marginally stable trajectories.  The associated configurations  are  quite similar to those of observed elliptical galaxies and  simulated cosmological halos,
and are necessarily different from the systems dominated by isothermal cores,  expected from entropy maximization in the context of the standard theory of violent relaxation.
It is suggested that this is the case because a system starting far from equilibrium does not reach a 'most probable state' via violent relaxation, but that this process comes to an end as the system finds and (settles in) a configuration where it can most efficiently wash out perturbations. We explicitly test this interpretation by means of direct simulations.
\end{abstract}

\maketitle

\section{Introduction}

  In a contemporary context, the Newtonian dynamics of a large number of self-gravitating particles describes 
the dynamics of putative dark matter, and to first order the formation and evolution of galaxies and clusters. Dark matter halos identified in cosmological simulations have similar (and  nearly) universal profiles, independent of the cosmological parameters or the masses of the halos or the epoch of their identification. These profiles have essentially the same form as those fitting early type galaxies (e.g., Navarro et. al. 1997;  Merritt et al. 2005). Yet there is no satisfactory theory predicting these preferred choices of dynamical 
equilibria.

Steady states are characterized by the absence of coherent collective modes, which
decay due to differences  between motions on neighboring particle trajectories and interactions between the coherent modes and 
individual particle motions. 
These  effects represent generic mechanisms driving gravitational systems toward collisionaless 
equilibria, ending the initial phase of violent relaxation, with its singnificant potential fluctuations 
and accompanying changes in particle energies.   The associated processes are often referred to as phase mixing and Landau damping 
respectively (Binney \& Tremaine 2008, hereafter referred to as BT) 
and should be most efficient when nearby trajectories rapidly diverge and spread in phase space.
However, attempts at relating these trajectory stability properties to the 
choice of $N$-body equilibrium --- or extracting virtually any interesting information at all from them --- are frustrated by the following phenomenon.  

As $N \rightarrow \infty$,  particles are essentially affected by the 
mean field potential corresponding to the smoothed out mass distribution
(indeed, it can be rigorously proven that the mean field
dynamics should describe asymptotically the time evolution of $N$-body 
system trajectories as $N \rightarrow \infty$; Braunn \& Hepp 1977).
The $N$-body problem is then transformed to $N$ 1-body problems
in this self consistent field, and steady state systems with sufficient symmetry are 
near integrable --- e.g., spherical systems are completely integrable when in a steady state,
supporting only regular orbits, which conserve as many isolating 
independent integrals of motion as degrees of freedom~\footnote{This assumes that the steady state 
is stable, in the sense that there are no collective instabilities that can drive the system away from equilibrium. If this is not the 
case, due to time dependent evolution, particle trajectories may in principle exhibit exponential divergence (and are thus not 'regular'), 
at least until a new equilibrium is found. Examples of such situations include spherical systems starting from initial conditions 
leading to radial orbit instability.} 

The phase space distances between trajectories moving on regular orbits generally diverges linearly in time~(BT). Thus the distance between trajectories of systems with only
regular orbits should also diverge linearly in time. 
This should imply that, for an $N$-body system with smoothed out potentials supporting 
only regular orbits, any exponential divergence timescale (indicating irregular trajectories and chaotic dynamics) should tend to infinity as $N$ is increased. Nevertheless, paradoxically, it has long been realized that point particle $N$-body systems invariably 
reveal exponential divergences on a timescale that does not
increase with $N$, even for systems with smoothed background potentials 
that only support regular orbits (Miller 1964; Goodman, Heggie \& Hut 1993; Kandrup \& Sideris 2001;  Hemsendorf \& Merritt 2002). 

This 'Miller instability' however saturates at progressively smaller scales as $N$ increases, and trajectories moving within a distribution of fixed point particles have been found to behave more and more like smooth potential counterparts as the number of background particles increased 
(Valluri \& Merritt 2000; Kandrup \& Sideris 2001; Hut 
\& Heggie 2002). This suggests that the Miller instability is a direct consequence of representation in terms of  point particles and that it is of little relevance if one is concerned with the gross structure and macroscopic evolution of large-$N$ gravitational systems. 

Furthermore, when
the singularity in the potential is removed (e.g. when $\phi \sim1/r \rightarrow  \phi \sim 1/\sqrt{r^2+  {\rm const}}$), the exponential divergence timescale between neighboring trajectories does increase with $N$ 
(Goodman, Heggie \& Hut 1993) and correlate with the evolution of dynamical variables like angular momentum (El-Zant 2002).  
It also worth noting that rigorous derivation 
of the approach toward the mean field limit in large-$N$ gravitational systems 
requires a smoothed two body potential (more precisely, $\phi$ is required to 
have continuous and bounded first and second derivatives; Braun \& Hepp 1977). This again suggests that, if progress is to be made in understanding the 
the evolution of macroscopic quantities (e.g., density and velocity dispersion profiles) in terms of the dynamical stability of trajectories, one has to move beyond the anomalous case of discrete point particles.  

The purpose of the present study is to show that, when the 
singularity in the two-body potential is removed, 
physically interesting information, pertaining to the
choice of equilibrium, can be extracted from the stability
properties of $N$-body trajectories. This will be done by using a 
geometric method for characterizing stability (Section~2), where the 
incompleteness of the pure point particle description  
of trajectory stability is clearest. When the singularity in the two body potential 
is smoothed the Miller instability is removed. What transpire instead are 
relations between macroscopic quantities characterizing equilibrium
configurations (as described in Section~3).  In Section~4,
we propose an interpretation of the approach to equilibrium and associated damping 
of coherent modes in terms of the stability properties of a system's trajectories; 
in Section~5  this interpretation
is tested and discussed. An explanation of the origin of equilibrium (spatial and phase space) density profiles of dark matter halos and elliptical galaxies follows from 
the context thus set.

\section{Geometric representation}

  The divergence of nearby trajectories can be also be deduced 
from  geometric representations of  motion and stability in 
gravitational systems 
(e.g., Gurzadyan \& Savvidy 1986; Kandrup 1990; Cerruti-Sola 
\& Pettini 1995; El-Zant 1997; El-Zant 1998). 
Though the geometric description is not unique (e.g., Casetti et. al. 2000), one well known
formulation has the advantage of involving a positive definite metric 
defined directly on the Lagrangian configuration manifold. For free 
particles, this manifold is simply the Euclidean space 
spanned by the $3N$ particle coordinates. In general, a system will 
move on a curved subspace, with velocity vectors defining its tangent space
(the phase space being the cotangent bundle; Arnold 1989). 
A locally Euclidean structure is generated on this space 
by the quadratic (in the coordinate derivatives) 
form of the kinetic energy (Arnold 1989). For a system 
conserving its total energy $E$, the associated conformally flat metric 
can be expressed in terms of the Cartesian  coordinates and the system kinetic energy 
$W = T(V) =  E - V$  as (e.g., Lanczos 1986) 
\begin{equation}
ds^{2}= W \sum_{3N} \left(dx^{\alpha} \right)^{2} \label{eq:L2}.
\end{equation}
The curvatures associated with this metric determine the stability of its trajectories. 
At the simplest intuitive level, it is clear why curvature should be related to stability.
For example, the distance between neighboring geodesics (great circles) on a sphere oscillates sinusoidally as one moves from pole to pole, while geodesics on a negatively curved saddle are hyperbolic and diverge away from each other. The case of a flat plane corresponds to marginally stability, with distance between straight lines increasing linearly along their lengths. 

   In the present case, the variation of the normal distance between neighboring geodesics (and therefore dynamical trajectories) defined by the above metric is determined by the two dimensional sectional curvatures $k_{\bf u,n}$ in $3 N$-dimensional configuration space, where ${\bf u}$ is the system velocity vector and 
${\bf n}$ define directions normal to it. And it can be rigorously shown that
when  all the $k_{\bf u,n}$ are negative, trajectories are exponentially unstable~(Anosov 1967; Arnold 1989). 
Isotropic $N$-body systems with singular two-body potentials are predicted to tend toward this 
state as $N$ is increased and direct two-body collisions are 
ignored~(Gurzadyan \& Savvidy 1986; Kandrup 1990).  
Perplexing trends, similar to those obtained from direct calculations (i.e.  
exponential divergence timescales that do not increase with $N$),  are then recovered when the full gravitational 
field is taken into account~(Kandrup 1990). This is a direct consequence of representing the matter distribution in terms of discrete point particles. In this paper we wish to move beyond this anomalous case.

We will be interested in how spherical $N$-body systems with isotropic velocities  damp out fluctuations, we therefore  
seek an averaged measure of  their response to random perturbations. Averaging the two dimensional 
curvatures over all possible directions ${\bf n_\mu}$ and trajectories (geodesics) at a given point in 
configuration space one obtains the 
the Ricci scalar; which, for $N \gg 1$, can be expressed as       
\begin{equation}
R=\sum_{\bf u,n} k_{\bf u,n} = - 3 N~\frac{\nabla^{2}W}{W^{2}} 
-  9 N^2~\frac{\parallel \nabla W \parallel^{2}}{4 W^{3}},
\label{scalar}
\end{equation}
where $\nabla W$ and $\nabla^{2} W$ are the gradient and Laplacian,  
taken with respect to coordinates $q_i = \sqrt{m_p}  x_i$, and
$m_p$  are their particle masses  (which will be assumed to be equal). 

The complete equations for trajectory stability will involve the two dimensional curvatures, and 
are usually written in terms of the geodesic distance $s$ (e.g., Kandrup 1990 for a clear derivation). 
A correspondingly transformned equation measuring average stability --- involving the scalar curvature and 
describing stability in the time domain --- 
is given by~(e.g., Cerruti-Sola \& Pettini 1995)
\begin{equation}
\frac{d^2 Y}{dt^2} + Q(t) Y = 0.
\label{eq:stab}
\end{equation}
Here $Y$  measures the mean divergence of trajectories and 
\begin{equation}
Q = \frac{2 W^2}{9N^2}~R - \frac{1}{4} \left(\frac{\dot{W}}{W}\right)^2 
+ \frac{1}{2} \frac{d}{dt} \left(\frac{\dot{W}}{W}\right).
\label{eq:flucs}
\end{equation}
Near dynamical equilibrium, $\dot{W} \rightarrow 0$ and the dominant term on
the right hand side will be the curvature term. 
For singular Newtonian potentials, 
Eq.~(\ref{scalar}) predicts negative $R$ (since the first term vanishes), and 
exponential instability is present even in the case of spherical equilibria. 
However, because of the singularities present when any two particle trajectories cross, 
one is here dealing with an incomplete manifold; 
it is not clear if straightforward
application of the geometric approach remains appropriate in this case~(Abraham \& Marsden 1978). The Miller instability thus figures again as a direct consequence of discrete representation in terms of point particles. As it will now be show however, the predictions are qualitatively different, and much more interesting, when one moves beyond this particular case.

\section{Equilibrium and pressure support}

When the singularity in the potential is softened, and as $N \rightarrow \infty$, the first term in Eq.~(\ref{scalar})
represents (through the Poisson equation) a  smoothed out density field integrated over the particle distribution
(with the softening length acting as smoothing parameter). In dynamical equilibrium, the two
terms on the right hand side of Eq.~(\ref{scalar}) are then comparable, reflecting a kinetic pressure versus 
gravity balance. To see this, assume $R=0$. In this case, one finds that
\begin{equation}
4 \pi G~W~\sum_{1}^N \rho_i = \frac{3 }{4}~ N ~ m_p  \sum_{1}^N a_i^2,
\label{eq:press}
\end{equation}
where $\rho_i$ and $a_i$ refer to the density and acceleration at the position of particle $i$.   
This reflects a balance between a pressure 
term (a density multiplied by the velocities squared), and a gravitational binding term; it 
is therefore expected to be at least approximately satisfied for systems in dynamical 
equilibrium. In fact, by dividing by $N$ and defining $\langle a^2  \rangle = \sum a_i^2/N$, $\langle \rho \rangle 
= \sum  \rho_i /N$,  $\langle v^2 \rangle$ as the means square speed of particles  
and $M = N m_p$, we can rewrite the above equation as   
\begin{equation}
W = \frac{1}{2} M \langle v^2 \rangle = 
 \frac{3 M~}{16 \pi G}  \frac{\langle a^2 \rangle}{\langle \rho \rangle}.
\label{eq:equ}
\end{equation}
Assuming that, in terms of a system characteristic length scale $r_s$, we can write  
$\langle a^2 \rangle \sim G^2 M^2 /r_s^4$ and $\langle \rho \rangle \sim M/r_s^3$, then 
Eq.~(\ref{eq:equ}) simply implies that 
$W \sim G M^2/r_s$  and $\langle v^2 \rangle  \sim G M /r_s$, 
which is characteristic of systems in dynamical equilibrium.

\section{Approach to equilibrium: an interpretation}

The results obtained thus far suggest that the two terms of  Eq~(\ref{scalar}) should be of 
the same order in the case of systems in dynamical  equilibrium. It is therefore reasonable to expect that such configurations  will correspond to  small values of $|R|$. Nevertheless, this does not imply that $R$ will exactly be zero for all such systems.  In particular, as shown in Appendix~A, $R$ will not 
vanish for systems dominated by nearly constant density isothermal cores, which are the natural product of the  process of classic violent relaxation gone to completion;; Lynden-Bell 1967; Nakamura 2000). However, it has been known for sometime now that such distributions 
do not match those of elliptical galaxies or dark matter halos. Instead, 
of all possible collisionless near-spherical equilibria, these structures exhibit 
a narrow range  of density profiles, characterized by central cusps that can be reasonably described by single power laws and 
extended halos with steeper power laws. It is therefore natural in our context 
to ask whether the latter correspond to $R \rightarrow 0$; and whether, in general, this condition gives 
us any insight into choice of gravitational equilibriia resulting from violent relaxation. 
We tackle the second question first, leaving the first 
one to the next section.

In this regard we recall that $R$ describes the stability characteristics of trajectories (through Eqs.~3 and 4). In dynamical equilibrium trajectories are stable if $R$ is positive and diverge exponentially if 
$R$ is negative. Before equilibrium is reached however the density and potential of a system 
started far from equilibrium will fluctuate. These fluctuations  can lead to trajectory instability even if the average $R$ is positive. The associated divergence and phase space mixing should result in the damping of coherent modes; and large scale fluctuations can only last up to and untill till these collective 
motions are damped. This means that the system should settle in a collisionless steady state when it finds a configuration where its response to fluctuations is least coherent; and this should correspond to dynamical equilibria where the trajectory divergence and phase mixing in response to fluctuations are maximal. At the same time, it is required that, in the absence of fluctuations, trajectories of
a spherical system should not be exponentially unstable in the infinite-$N$ limit, so as to recover an integrable system. 
This implies that $R$ cannot be negative when the system is in exact equilibrium.

Eq.~(\ref{eq:stab}) suggests that configurations that satisfy these requirements have
$R \rightarrow 0$ at equilibrium. This is easy to understand qualitatively: if average $R$ is positive and large, fluctuations (in $R$ and $W$)
have a smaller 
effect, and solutions of (\ref{eq:stab})
tend to those of a harmonic oscillator; conversely, 
if the average $R$ is near zero, $Q$ is dominated by fluctuations, which can 
lead to significant instability, especially that in that case fluctuations in $Q$ can lead to negative values. 
A system starting far from equilibrium, with initial conditions that allow for violent relaxation, should
then end up in such marginally stable states.

This intuitive argument is investigated more detail in Appendix~B for the case of periodic oscillations and stochastic fluctuations. In the next section numerical simulations will be presented that suggest that systems starting further from equilibrium do indeed end up with $R$ closer to zero.  But before closing this section it is important to note that though the above interpretation stands in seeming contradiction with that proposed by Kandrup and collaborators, whereby it was suggested that trajectory divergence causes relaxation rather than subduing it, the clash here is only partial. The aforementioned authors reached their conclusions while studying trajectories moving through either stationary non-integrable 
potentials (e.g. Kandrup \& Siopis 2003), or in particular spherical potentials with time dependent mass or length scales (but which were  otherwise rigid and fixed; e.g Kandrup et. al. 2003). In the latter case they found that the time dependence can be a significant source of trajectory divergence, which may drive phase space mixing associated with 
violent relaxation. Up to this point there is no contradiction with the interpretation outlined above; 
fluctuations in the density and potential will indeed lead to trajectory instability and phase space mixing. However, when one takes into account the evolution of the self consistent field, it will be 
realized that the trajectory divergence and phase space mixing should also lead to the decay of the  
very collective modes that are causing the time dependence (and associated trajectory instability, mixing and relaxation) in the first place. When these are efficiently damped, no further evolution can occur (In the collisionless limit). This is in line with our contention that the final product of violent relaxation do not correspond to classic 'most probable' configurations but to (quite different) ones, which are most efficient in damping out collective modes via trajectory instability and associated phase space mixing.

\section{Testing the interpretation}

   We now examine the above interpretation 
by means of direct $N$-body simulations. Since we are interested
in estimating the densities and accelerations in the continuum limit, 
we use a technique~(Hernquist \& Ostriker 1992),
which expands  the density and potential in smooth functional series. In this study, we  
are interested in strictly spherical configurations and the 
results described below only make use of the radial expansions in this scheme, which 
is carried up to order $30$ (however it has been verified that the results are not significantly altered if
azimuthal expansion terms are included, as noted in the discussion below).
 Systems are sampled using a $100 000$ particles (except for one case with 500 000 as also described below), and  
started from homogeneous spatial initial 
conditions inside a unit sphere of unit mass. This configuration has  
a natural timescale $\tau_D  = (G \rho (0))^{-1/2} = \sqrt{4 \pi / 3}$, which
will be our time unit.

We have run configurations starting with  isotropic velocities that are either constant 
or following various decreasing or increasing functions of radius. 
The results shown here correspond to the latter case, for the following reason.
Unless the initial velocities are vanishingly small, uniform initial velocities will
imply that the initial 'pseudo-phase space density' $\rho_p = \rho/\sigma^3$, does not 
diverge anywhere. On the other hand, collisionless evolution implies that 
$\rho_p$  generally 
decreases along particle trajectories~(BT), so that the maximal final $\rho_p$  cannot exceed the initial one. One of our aims is to explain the dynamical origin of     
cosmological halos, which have  (a centrally divergent) 
$\rho_p \sim r^{-1.875}$~(Taylor \& Navarro 2001), which in turn requires an intial phase space density that is also divergent (so that this initial density can, in principle, be mapped via collisionless dynamics into that characterizing dark halos.
From among various functional forms satisfying this condition that were tried, results will be shown for initial $\sigma \sim r$, noting that the trends reported seem generic.

\subsection{The approach toward marginal stability}

We vary the initial virial ratio ${\rm Vir} = 2 W/ |V|$ from $1$, corresponding to near 
equilibrium, to $0.125$, corresponding to a system started far from equilibrium.
Given the sampling errors for finite $N$, $R$
will never tend to zero exactly in practice; and the absolute values of the 
densities and accelerations (and local dynamical times) strongly 
depend on the central concentrations of the final configurations. 
A relative measure is thus required for comparing these. We use the normalized 
quantity 
\begin{equation}
R_n = 
\left(4 \pi G W \sum_1^N \rho_i\right) \left(\frac{3 }{4}~ N ~ m_p  \sum_1^N a_i^2 \right)^{-1}
 - 1,  
\label{Rn}
\end{equation}
which  measures the departure from equality in Eq.~(5).

\begin{figure}[t]
\includegraphics[angle=-90,scale=.65]{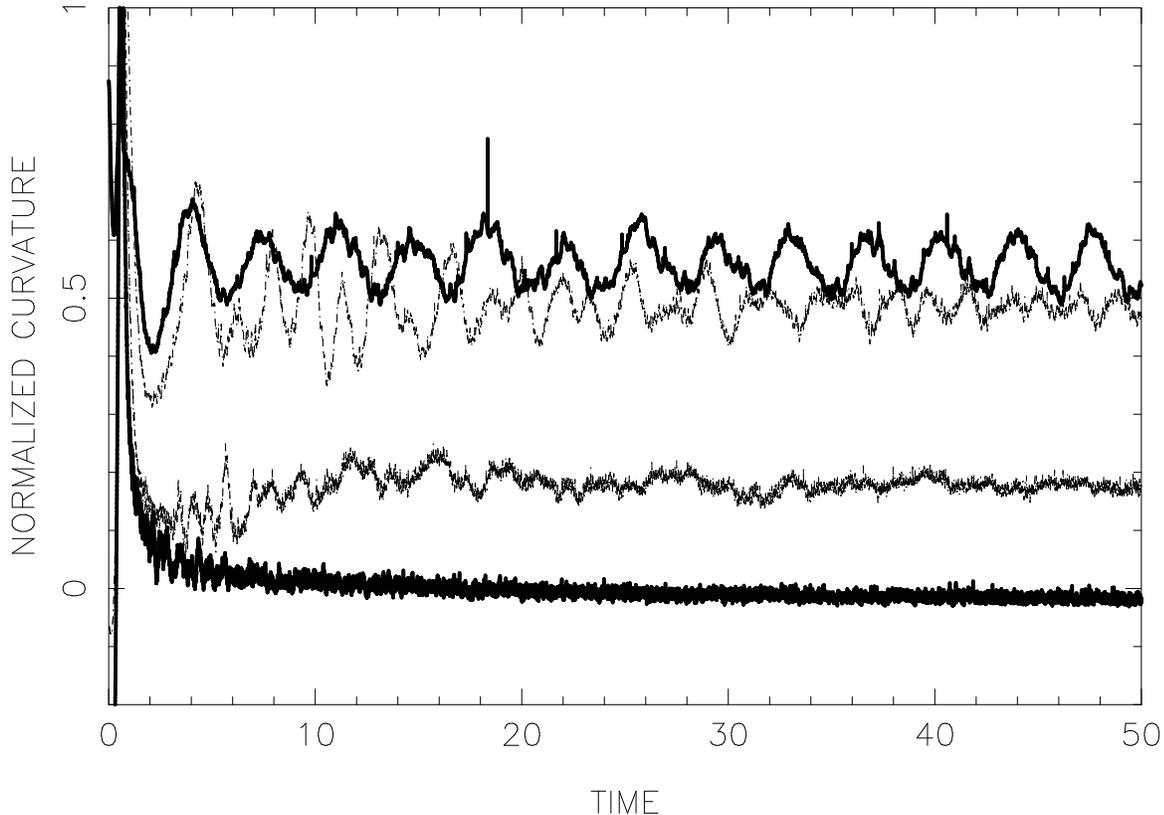}
\caption{Time evolution of normalized curvature $R_n$ (Eq.~\ref{Rn}), measured 
in  initial dynamical times, for systems starting with (from top to bottom) 
virial ratios, 1, 0.5, 0.25 and 0.125} 
\label{curvatures}
\end{figure}
\begin{figure}[t]
\includegraphics[angle=-90,scale=.65]{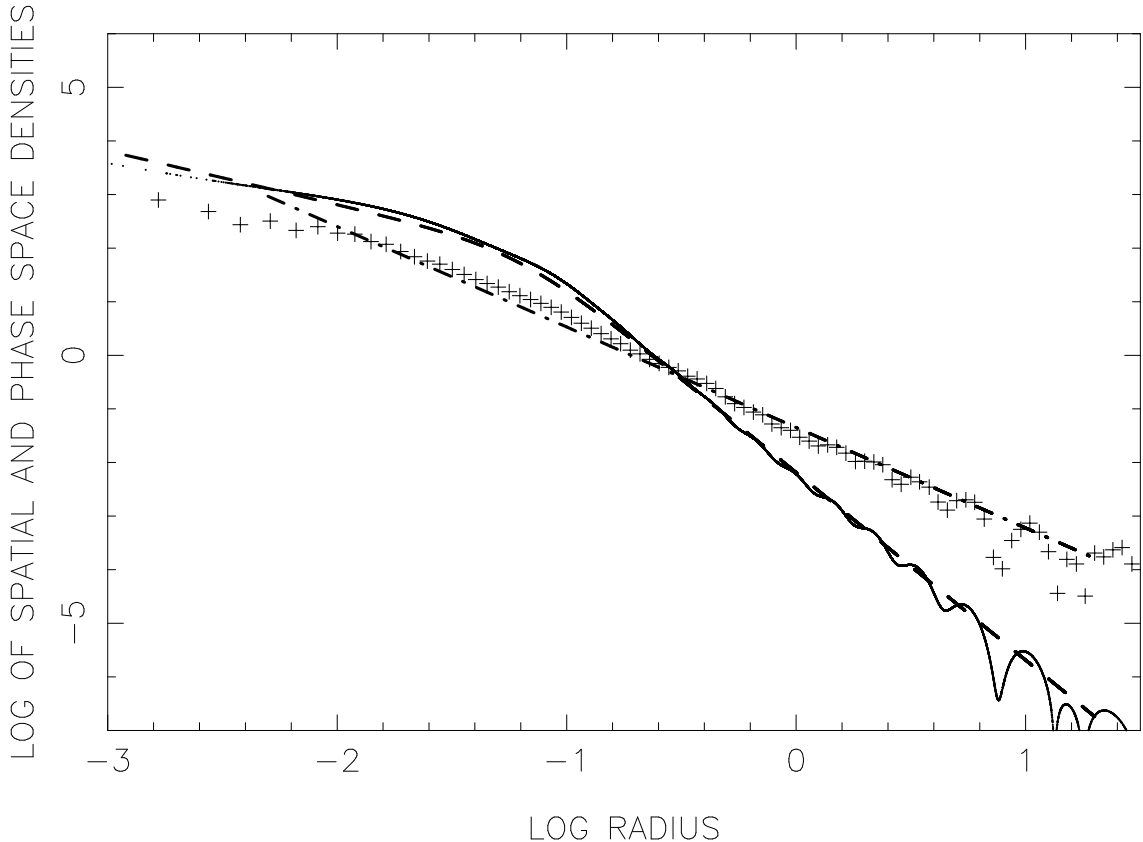}
\caption{Equilibrium density (solid line) and phase space density
(crosses) of system starting with virial ratio 0.125. Dashed lines
are eye fits with$\rho \sim 1/r (1+r^{2.5})$ and $\rho_p \sim r^{-1.875}$} 
\label{densfit}
\end{figure}

  The results are in Fig.~1, which shows that the  equilibrium $R_n$ is indeed always of order unity or 
less, and that the further from equilibrium the initial state, the closer the evolved system 
is to a configuration that minimizes $|R_n|$, in line with the general interpretation presented in the previous section. The initial $R_n$ is zero when ${\rm Vir} =0.5$;
systems  with initial ${\rm Vir}< 0.5$ start 
with $R_n < 0$ and then explore state space until they achieve configurations where (according to Eq. ~\ref{eq:stab})oscillations are efficiently damped.

We note that a similar trend was found when systems were started with constant initial phase space density
$\rho_p$, except that generally larger final values of $R_n$ are obtained. 

\subsection{Interpreting the dark matter halos and elliptical galaxies}

The origin of the density profiles of bound structures identified in the context
of cosmological simulations have been intensely investigated; the mechanism
for their advent was sought for example 
in terms of the merging histories of smaller structures in the way of forming larger 'halos'
(e.g., Syer \& White 1998; Dekel et. al. 2003). However, the same  sort of
structures tend to materialize even in simulations where a cutoff has been applied to the power spectrum and the collapse is monolithic (e.g., Huss et. al. 1999; Moore et. al. 1999; Wang \& White 2009). These profiles thus seem to arise nearly independently of the condition of their formation; merging may conserve their form (e.g., Boylan-Kolchin \& Ma 2004;
El-Zant 2008),  but is not necessary to its genesis.

It is thus natural to inquire if a simple explanation may be sought in terms 
of the stability issues we have addressing in this paper. We know from the previous 
subsection that systems started far from equilibrium to undergo simple monolythic  collapse
have near-zero final average curvatures; they are therefore marginally stable according to the 
interpretation outlined in Section~4. If, as it is suggsted,  violent relaxation should generically terminate in such states, and if 
violent relaxation is predominatly responsible for the structure of dark halos, then those halos 
should correspond to configurations with near-zero curvature.  This would be the case if they  
corresponded to the final configurations of our simple collapse simulations --- despite the 
lack of cosmological context or associated merging processes in these simulations. 

Indeed, one finds, as shown in Fig.~2, that the equilibrium density and  phase space density profiles of  systems started far from equilibrium 
and undergoing simple monolithic collapse, starting from 'cold' initial conditions, 
are remarkably similar to those of cosmological halos (and also shallow cusp elliptical galaxies; Merritt et al. 2005),
despite the lack of cosmological context.  
Systems started from virial equilibrium on the other hand 
tended to end a up with large central cores (as do systems started with initial phase space 
density profiles that disallow the final configurations shown).  This simulation was repeated  with 
several values of  radial and azimuthal expansion orders in the Herquist-Ostriker scheme, including one simulation with
$500 000$ particles and 20 radial expansion  terms and 12 azimuthal ones, without significant change in the 
results.  

There are some difference between the configuration shown and idealized halo profiles. 
Our outer density profile is steeper than the $1/r^3$ form of
cosmological halos; however finite configurations cannot have  $1/r^3$ outer  profiles. (Cosmological halos are not isolated equilibrium structures as their outer regions are subject to mass infall, which reduces the steeness of the outer profile; e.g., Gott 1975).
The functional form of our empirical spatial density fit is also slightly different from
standard 'NFW'  (where $\rho_{\rm NFW} \sim \frac{1}{r (1 + r)^2)}$), and there seems to
be some flattening toward the center of the $1/r$ cusp. Yet, similar departures from ideal NFW often also apply to cosmological halos (Navarro et. al. 2004).

   Our argument for the production of the 'universal' profiles --- when the initial phase space density 
allows for their advent --- is that the associated configurations are most efficient in
damping out large scale coherent oscillations. We now explicitly verify that
among the configurations we have simulated (which always have positive or near zero 
average curvature once they reach a quasi-steady state), those with smaller $R_n$  do
efficiently damp out fluctuations. This
is already  apparent in Fig.~1, where fluctuations are quickly damped for 
such systems. This may be intuitively expected, for these systems lack
homogeneous harmonic cores with nearly constant orbital frequencies and coherent response, 
and similar trends were indeed observed previously (e.g., David \& Theuns 1989). 
Indeed Eq.~(3) allows in principle for the existence of long lived coherent modes 
when the average $R$ is large (as discussed in Appendix~B).

Further verification
is obtained by perturbing the final states and measuring
variations in the density distributions as  
systems, so perturbed, are evolved. The results 
of one such experiment, where the spatial coordinates were decreased by $10 \%$,
are shown in Fig.~3. There, the system that was perturbed from a state with smaller 
$|R_n|$  is seen to quickly settle back to an equilibrium that is closer to the unperturbed one, 
and to gradually close in further on it (as the outer regions, with large crossing times, evolve.

From the above, one can conjure that the observed robustness of the universal profiles when perturbed may also be a reason behind their conservation under processes involving significant potential fluctuations
--- such as mergers, accretions and fly bys ---  that are important in the context of cosmological 
simulations.

\begin{figure}[t]
\includegraphics[angle=-90,scale=.65]{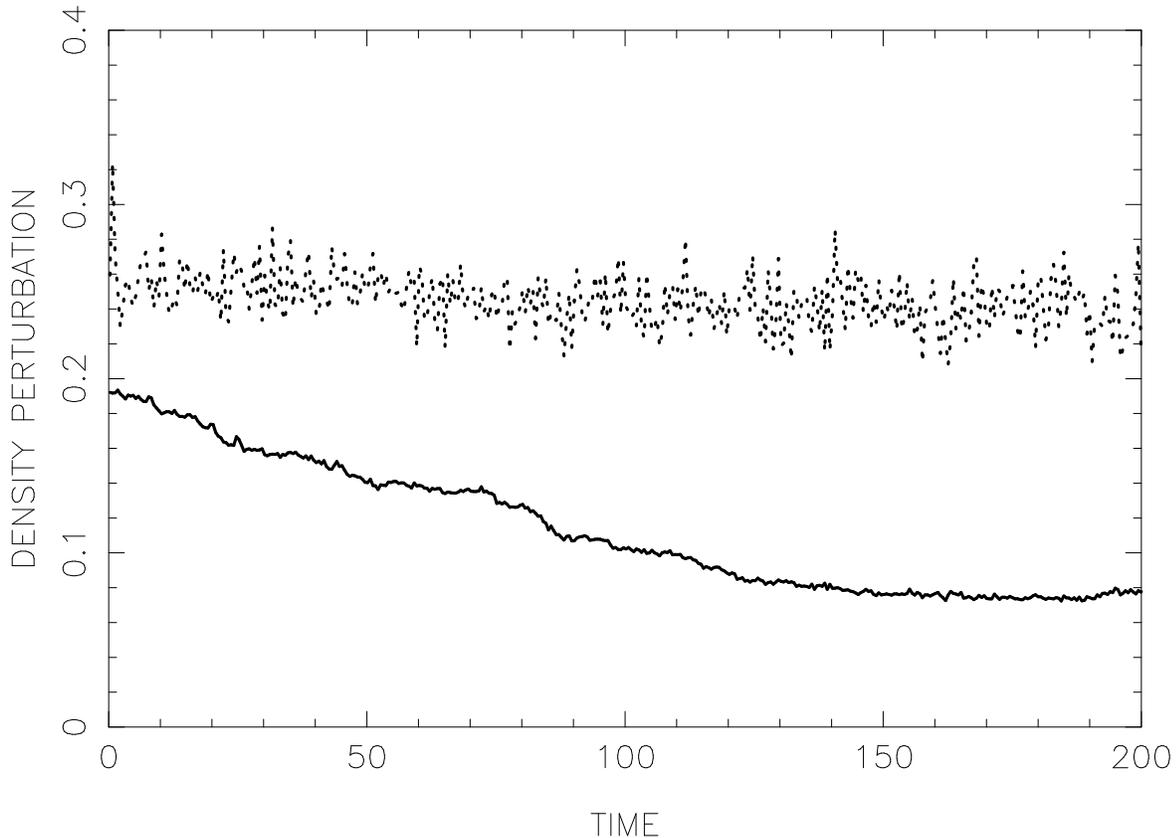}
\caption{Evolution of density perturbations, defined as $\sum_{i=1}^{N} (\rho_i (t = 0)  -  \rho_i (t))^2 /\sum_{i=1}^{N} \rho_i^2 (t = 0)$, when coordinates of final states of systems  starting from  virial ratio 0.125 (solid line)  and  1 (dotted line) are perturbed by $10 \%$} 
\label{perturb}
\end{figure}


Finally we note that, despite the formal similarity between profiles of elliptical galaxies and dark halos,  
the above interpretation strictly speaking applies only to the advent of dark halo configurations. 
The case of early type galaxies is much more complicated. The currently favored theory for the formation of early 
type galaxies involves a 'two-phase' process: 
a highly dissipative initial stage accompanied by intense star formation,   
followed by a succession of dissipationless mergers (e.g., Oser et. al. 2010).   
The simple dissipationless collapse simulations presented here are evidently not of direct relevance
to this rather involved two-phase model. However, though the initial phase is 
highly dissipative, it sets the initial conditions for largely dissipationless dynamics, presumably
dominated by significant fluctuations characterizing an evolving time dependent system (this should be especially the case if star formation is sufficiently efficient and the associated burst sufficiently intense; indeed, Oser et. al. suggest this phase 'bears an uncanny resemblance' to claasic monolythic collapse models). If the general interpretation (of Section~4) is correct, marginally stable systems should in principle be generic products of such a situation. Furthermore, as noted above, resulting profiles should be conserved in the second phase (of dissapationless merging).  This seems plausible; however it is evidently beyond the scope of this paper to explicitly test the interpretation presented here in the complex context of the two-phase theory of the formation of early type galaxies.   

\subsection{A note on discreteness noise and relaxation timescales}

   Our main concern in this paper has been with 
the way mean field fluctuations are damped as a result 
of the  mixing that occurs due to the instability 
of phase space trajectories. Historically, however, it has been
(controversially) argued that when trajectory instability stems 
not from large scale fluctuations but from discreteness noise, 
a new mode of relaxation, different from standard two body relaxation and 
with a correspondingly different timescale, arises (Gurzadyan \& Savvidy 1986).

   As noted in the introduction and in Section~2, when the full self consistent 
gravitational field is taken into account, and with a singular potential,
the prediction is that system trajectories will always be unstable 
on the timescale of a dynamical time or less  (the Miller instability). Gurzadyan \& Savvidy on the other hand
employed a Holtsmark distribution (e.g., Chandrasekhar \& von Neumann 1942), 
where the force function is not self consistent but is dominated 
by nearest neighbor interactions (and thus varies as~$1/\sim N^{1/3}$). 
They thence deduced that the instability timescale scales 
varies as $\sim N^{1/3} \tau_D$, and identified 
this with a new 'collective' relaxation time. 

   Three things are to be noted in our present context. First, assuming
that that fluctuations due to discreteness noise 
(in the curvature and associated quantities, as density and force) scale 
as $\sim 1/\sqrt{N}$, then Eqs.~(\ref{eq:stab}) and (A3) predict that, 
when the curvature is much larger than 
discreteness noise fluctuations, unstable trajectories would diverge on 
a timescale $\tau_e \sim N \tau_D$; while $\tau_e \sim N^{1/3} \tau_D$ 
when the reverse is true. 
Thus, for statistically significant $N$ and
unless the curvature is extremely closely fine tuned to zero,
the divergence timescale for nearby trajectories will scale more as
the standard two body relaxation time than the Gurzadyan \& Savvidy
timescale. Secondly, even with an $N^{1/3}$ scaling, the source of the 
instability in our case are fluctuations around a near-zero curvature, calculated 
taking into account the full self consistent field,and not a 
small negative curvature arising from closest neighbor forces,
as in the calculation of Gurzadyan \& Savvidy.  

 The third point concerns interpretation. Relaxation on the two body 
timescale, as usually understood in stellar dynamics, is accompanied by 
structural changes in the collisionless equilibrium configuration of the system,
due to diffusive random walks undergone by the integrals of motion 
associated with the mean field potential; e.g., energy
relaxation leads to core contraction and collapse. Whether
divergence between neighboring trajectories directly leads to such effects on 
the corresponding exponential timescale is not clear,
since such divergence will include changes in phases between particle trajectories as well vartiations in the integrals of motion associated with the corresponding trajectories in the mean field potential. Much further detailed investigation would be needed to establish a precise relation (if any) with 
relaxation as normally understood in stellar dynamics. 

On the other hand, it seems clear that the efficiency of phase space mixing due to 
exponential divergence should  directly correlate with the decay of coherent collective modes, 
as even simple phase mixing (without any exponential instability or changes in the integrals of motion) is effective in damping of these modes. Thus the interpretation of the effect of 
trajectory  divergence in the context of the present paper appears to 
rest on firmer ground. An analogous interpretation in the case of discreteness 
noise is that the timescale of decoherence of collective modes 
should depend on the number of particles in the system (as observed, e.g., 
by David \& Theuns 1989). 

It is hoped that the present exploratory study has shown that the methods employed, which 
have hitherto been of little use, are potentially powerful tools in 
understanding the structure of gravitational systems; and that it would
stimulate further work on these fundamental --- yet lately little researched --- 
problems pertaining to the evolution of Newtonian gravitational systems.

\section{Conclusion}

According to the classic idea of violent relaxation, a collisionless system starting  far from dynamical equilibrium reaches such a state  via collective oscillations. If the associated mean field fluctuations 
persist until the bulk of the system is described by a most probable phase space distribution, 
the resulting distribution functions are of Maxwellian form, which are quite different from the distribution functions of 
dark matter halos for example (e.g., Widrow 2000). The associated density profiles are dominated by isothermal cores (Lynden-Bell 1967), which are not observed 
in elliptical galaxies or in simulated cosmological halos. In contrast, in the scenario proposed here, violent 
relaxation should  persist only until a system, starting far from equilibrium, finds a  configuration where 
collective oscillations are efficiently damped; and such 
configurations  support  dynamical trajectories that are marginally stable. 
This condition necessarily leads to equilibria that 
are different from isothermal (except for exact singular isothermal pheres), 
and which are remarkably similar to those observed and modeled,  
provided that the initial conditions do not impose phase space constraints
that disallow evolution to such states. 
These systems do not 'ring' and are more robust; in the sense that, when perturbed, they 
rapidly return near their initial states, as opposed to configurations dominated by harmonic 
cores, where small perturbations give rise to significant structural 
changes accompanied by long-lived collective modes.   There is an  intuive explanation of this result; 
as  opposed to configurations dominated by nearly constant density harmonic 
cores, these systems display relatively large orbital frequency ranges, and  are thus more efficient 
in damping collective modes.  It is also possible that this is a reason behind the conservation of the universal profiles through major mergers and other processes involving significant potential fluctuations that are important in a cosmological context, though this contention has not been investigated explicitly here.

\acknowledgments

I would like to thank the referee for comments that lead to significant improvement in substance and
 presentation, 

\appendix
\section{The absence of marginally stable non-singular isothermal spheres}

Replacing $a_i$ and $\rho_i$ in Eq.~(\ref{eq:press}) by their continuum limit counterparts, 
and integrating over the spherical mass distribution, one gets 
\begin{equation}
4 \pi G W \int \rho^2 4 \pi r^2 dr =  \frac{3}{4}  M  \int  a^2 4 \pi \rho r^2 dr,
\end{equation}
where $\rho = \rho(r)$ is the radial density.
Defining the local acceleration $a (r) =  - G m /r^2$
(with $m$ the mass enclosed within radius $r$), one can write 
\begin{equation}
4 \pi W \int \rho^2 r^2 dr = \frac{3}{4}  M G \int \frac{m^2}{r^2} \rho dr. 
\end{equation} 
In dynamical equilibrium 
one can use the Jeans equation for an isotropic system, $\frac{d} {dr} \rho \sigma^2 = - G \rho \frac{m}{r^2}$ 
(e.g., BT) ,  to get
\begin{equation}
4 \pi W \int \rho^2 r^2 dr = -  \frac{3}{4}  M \int m~d (\rho \sigma^2), 
\label{curcrude}
\end{equation} 
where $\sigma^2 = \bar{v^2} (r)$ is the one-dimensional velocity dispersion.
There is one easily identifiable solution of Eq.~(\ref{curcrude}); it corresponds to  $\rho \sim 1/r^2$ and $\sigma = {\rm constant}$. This is a  singular isothermal sphere; however its gravitational 
force, and both terms in Eq.~(\ref{curcrude}), diverge as $r \rightarrow 0$.
   
Well behaved systems --- including those with inner power law density profiles shallower than 
$r^{-5/3}$, as is the case of simulated dark matter halos --- have  
$ m \rho \sigma^2=0$ as $r \rightarrow 0$ and $r \rightarrow  \infty$ (e.g., when an outer density 
profile is  
Integrating by parts, noting this, and  
that the kinetic energy is related to the average  velocity dispersion by $W = \frac{3}{2} M \langle \sigma^2 \rangle_r$, and $dm =4  \pi r^2  \rho dr$, one gets
\begin{equation}
\langle{\sigma^2} \rangle_r ~ \int \rho~dm = \frac{1}{2} \int \rho \sigma^2~dm.
\label{eq:equi}
\label{fund}
\end{equation}
If effective  pressure and temperature functions can be assigned, 
then $P = \rho \sigma^2$ and $W= \frac{3}{2} N k~ \langle T \rangle_r$, and 
\begin{equation}
\int P~dm = 2~\frac{k~\langle T \rangle_r }{m_p} \int \rho~dm. 
\end{equation}
Save for the integrations over the mass distribution, this form is that of an ideal gas equation of state.  
Nevertheless, because of the factor  $2$, there are no associated isothermal states.

Thus nearly isothermal systems 
(with non-diverging $m \rho \sigma^2$) cannot be marginally stable 
in the sense of $R \rightarrow 0$. Therefore they should not result from
violent relaxation according to the scenario proposed in this paper, even though thay are predicted via entropy maximization in the context of classic violent relaxation theory. Eq.~(\ref{eq:equi})  also shows that this is the case 
with any configuration dominated by a nearly constant density core. This is in line weith the numerical results of Section~5.

\section{Fluctuations and trajectory divergence}

Eq.~(\ref{eq:stab}) can be written in terms of the average $Q$
and  time varying term $Q_1$ as 
\begin{equation}
\frac{d^2 Y}{dt^2} + \left( \langle Q \rangle + Q_1 (t) \right)) Y = 0.
\label{eq:av}
\end{equation}
In general, phase space averages are not well defined for gravitational systems (which 
have a non-compact phase space). Nevertheless, every  exact  ($N \rightarrow \infty$)
collisionless equilibrium has a characteristic $R$  and associated  
$k_0 =  \frac{2 W^2  R}{9~ N^2}$. A system oscillating about such an equilibrium will have an 
average $Q$ close to this value (as the second and third terms in Eq.~4 are zero in exact equilibrium, their contributions to average $Q$ being second order in the fluctuations), with RMS fluctuations $\sigma^2_k = \langle Q^2 \rangle - \langle Q \rangle^2$.  If the time variations are periodic this is a Hill equation, with its well known instability strip increasing in measure as $\sigma_k$ increases relative to $k_0$, even for positive $k_0$ (e.g. Magnus \& Winkler 2004). Solutions will always be unstable if the curvature can reach negative values. On the other hand, for positive  $k_0$ that is large relative to the time varying term, the instability regions are vanishingly small, since n that case, solutions Eq.~(\ref{eq:av}) tend to those of a harmonic oscillator. These results suggest that for a given level of periodic fluctuations instability is more likely to occur (and it occurs with smaller divergence timescale), as the average curvature $k_0$ decreases. 

The existence of stable solutions for large $k_0$ may account for the apparently 'locked' long lived 
oscillatory modes observed in systems starting near virial equilibrium (Fig.~1 and Fig~3). As pointed out in the discussion of Fig.~3, there is also an intuitive explanation  of these oscillations: the associated  configurations exhibit 
dominant harmonic cores, where orbital frequencies change little, facilitating  the persistence of coherent modes.            
This result is in line with those of the previous appendix, showing that systems dominated by nearly constant density cores are unlikely to correspond to near zero curvature.

If the fluctuations in $Q$ 
can be considered as stochastic (more precisely as normally distributed 
random variables), one can use a standard formulation for solving stochastic 
equations, developed by van Kampen (1976), to show that solutions are always unstable, even for positive $k_0$, with the timescale of exponential divergence between nearby trajectories decreasing as the RMS fluctuations $\sigma_k$ increase relative to $k_0$. To see this, assume that the fluctuations $\sigma_k$  have characteristic  timescale $\tau$. This gives 
rise to the following  estimate of the exponential divergence timescale of the quantity $Y$
 (e.g., Eqs. 79 of~Casetti et. al. 2000)
\begin{equation}
\lambda = \frac{1}{2} \left(\Lambda - \frac{4 k_0}{3 \Lambda}\right),
\end{equation}
with
$\Lambda^3 = \sigma^2_k \tau + \sqrt{\left(\frac{4 k_0}{3}\right)^3 + \sigma^4_k \tau^2}$.

The natural timescale for fluctuations in a gravitational system is the dynamical (or crossing) time, so
$\tau = \tau_D$.  And since $Q$ comes in units of inverse time squared, we can write
$k_0 = a/\tau_D^2$ and $\sigma = b/\tau_D^2$, and express the exponentiation 
time associated with trajectory divergence in terms of the natural timescale as 
\begin{equation}
\frac{\tau_e}{\tau_D} =   \frac{6 \left(b^2 + \sqrt{\left( \frac{4}{3} a \right)^3 + b^4}\right)^{1/3}}
{3 \left(b^2 + \sqrt{\left( \frac{4}{3} a \right)^3 + b^4}\right)^{2/3} - 4 a}.
\label{petstab}
\end{equation}
This is proportional to  $a b^{-2}$ for $a \gg b$ and 
tends to  $2^{2/3} b^{-2/3}$ as $b \gg a$.
For  a given fluctuation level $b$, the decoherence of trajectories due to local divergence will 
be more efficient,  i.e. its exponential time will be smaller, for smaller $a$,
and therefore $k_0$ and average $R$.

\end{document}